\documentclass[letterpaper, 10 pt, conference, twocolumn]{ieeeconf}

\IEEEoverridecommandlockouts
\usepackage{cite}
\usepackage{amsmath,amssymb,amsfonts}
\usepackage{multirow}
\usepackage{algorithm}
\usepackage{algorithmic}
\usepackage{graphicx}
\usepackage{textcomp}

\usepackage{multirow}
\usepackage[table,xcdraw]{xcolor}

\usepackage{soul}
\usepackage{subcaption}
\usepackage[draft]{hyperref}
\def\BibTeX{{\rm B\kern-.05em{\sc i\kern-.025em b}\kern-.08em T\kern-.1667em\lower.7ex\hbox{E}\kern-.125emX}} % [RHT] Adding some color definitions

\definecolor{deep-red}{RGB}{192, 0, 0}
\definecolor{deep-purple}{RGB}{120, 0, 170}
\definecolor{good-green}{RGB}{0,175,0} 
\definecolor{purple}{RGB}{210, 0, 210}

\newcommand{\hi}[1]{\textcolor{black}{#1}}
\begin{document}

% Adjust spacing around floats (reduce it)
\setlength{\textfloatsep}{8pt plus 2pt minus 4pt}

\title{Improving Surgical Situational Awareness with Signed Distance Field:\\ A Pilot Study in Virtual Reality}
\author{%
     Hisashi Ishida$^{*1}$, Juan Antonio Barragan$^{*1}$, Adnan Munawar$^1$, Zhaoshuo Li$^1$,\\ \hi{Andy Ding$^2$}, Peter Kazanzides$^1$, Danielle Trakimas$^2$,\\ Francis X. Creighton$^2$, and Russell H. Taylor$^1$%
     \thanks{$^1$Department of Computer Science, Johns Hopkins University, Baltimore, MD 21218, USA. $^2$ Department of Otolaryngology-Head and Neck Surgery, Johns Hopkins University School of Medicine, Baltimore, MD 21287, USA. *These authors contributed equally. 
     Email: {\tt \{hishida3, jbarrag3\}@jhu.edu}}
     % {\tt \{hishida3, jbarrag3, zli122, pkaz, rht\}@jhu.edu, amunawa2@jh.edu, , misha@cs.jhu.edu, \{dtrakim1, francis.creighton\}@jhmi.edu}}
}

\maketitle

\begin{abstract}
The introduction of image-guided surgical navigation (IGSN) has greatly benefited technically demanding surgical procedures by providing real-time support and guidance to the surgeon during surgery. \hi{To develop effective IGSN, a careful selection of the surgical information and the medium to present this information to the surgeon is needed. However, this is not a trivial task due to the broad array of available options.} To address this problem, we have developed an open-source library that facilitates the development of multimodal navigation systems in a wide range of surgical procedures relying on medical imaging data. To provide guidance, our system calculates the minimum distance between the surgical instrument and the anatomy and then presents this information to the user through different mechanisms. The real-time performance of our approach is achieved by calculating Signed Distance Fields at initialization from segmented anatomical volumes. Using this framework, we developed a multimodal surgical navigation system to help surgeons navigate anatomical variability in a skull base surgery simulation environment. Three different feedback modalities were explored: visual, auditory, and haptic. To evaluate the proposed system, a pilot user study was conducted in which four clinicians performed mastoidectomy procedures with and without guidance. Each condition was assessed using objective performance and subjective workload metrics. This pilot user study showed improvements in procedural safety without additional time or workload. These results demonstrate our pipeline's successful use case in the context of mastoidectomy.

\end{abstract}

\section{Introduction}
Technically demanding surgical procedures, such as skull base surgical procedures, have greatly benefited from the introduction of image-guided surgical navigation (IGSN). IGSN systems use preoperative models of patient anatomy derived from Computed Tomography (CT) or Magnetic Resonance Imaging (MRI) images for surgical planning.  
Intraoperatively, these models can be registered to the actual patient, and the motions of surgical instruments relative to the patient can be tracked. Using this information, a navigation system can provide the surgeon with real-time support information and guidance, leading to improved surgical situational awareness, lower mental demands, and higher patient safety \cite{luz_impact_2016}. 

\begin{figure}
    \centering
    \includegraphics[width=0.5\textwidth]{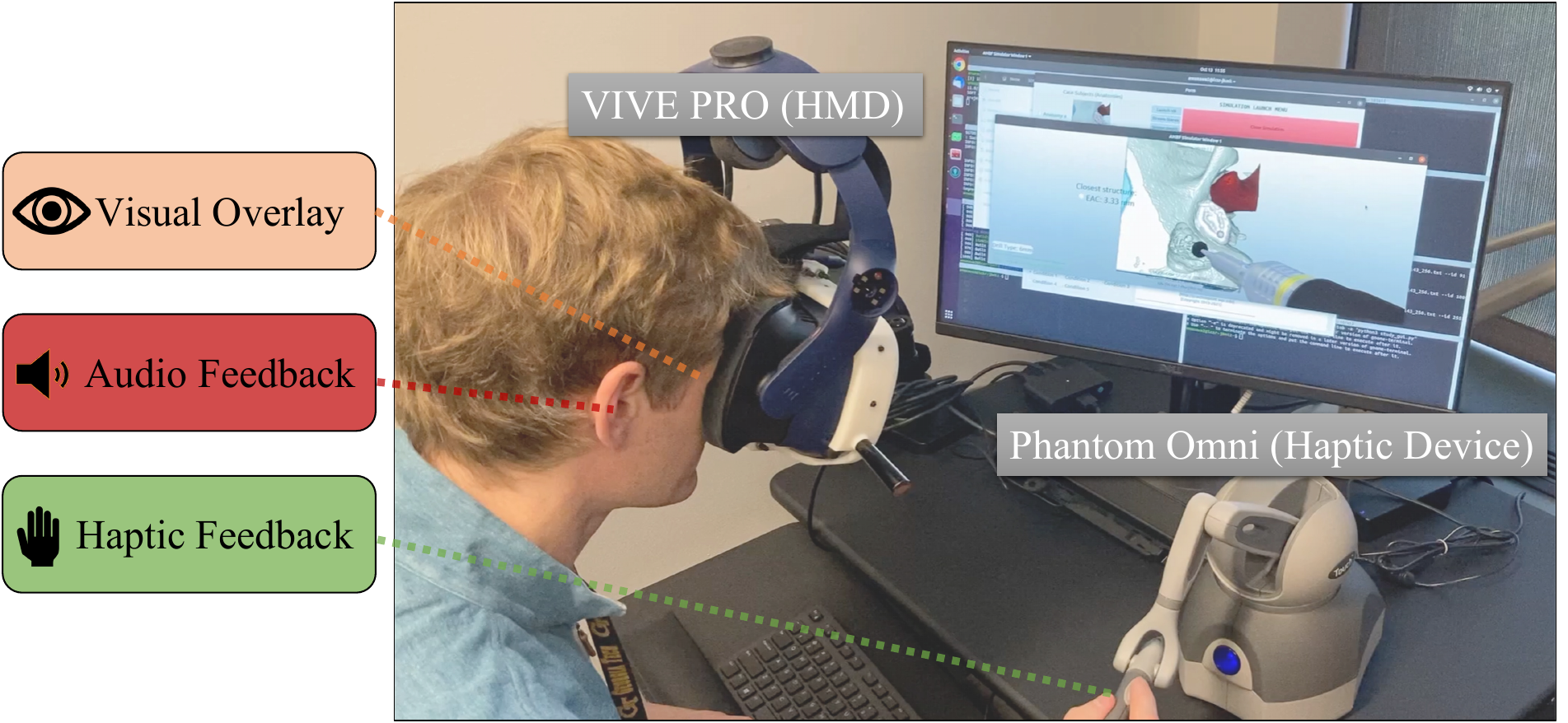}
    \caption{Hardware setup for virtual drilling simulator. The hardware setup emulates a real mastoidectomy environment with the head-mounted display (HMD) in place of a stereo microscope, the haptic device in place of the surgical drill, and a foot pedal interface for actuating the drill. The setup is housed on a movable cart for portability.}
    \label{fig: Exp_setup}
\end{figure}

The effectiveness of a navigation system depends not only on accurate registration algorithms but also on carefully tailoring the information presented to the surgeon \cite{matsumoto_augmented_2020}. IGSN systems can be broadly categorized into systems that only provide support information to the surgeon and systems that directly affect the surgeons' actions, e.g., a robotic surgical system that enforces safety barriers \cite{mingli2007vf,li2020anatomical}. IGSN can be further categorized depending on the medium used to provide navigational information, e.g., visual, auditory, and tactile information. Given the broad spectrum of available options to present feedback, identifying optimal modalities for a specific surgical context is challenging. 

\begin{figure*}[htp]
      \centering
      \includegraphics[width=1.0\textwidth]{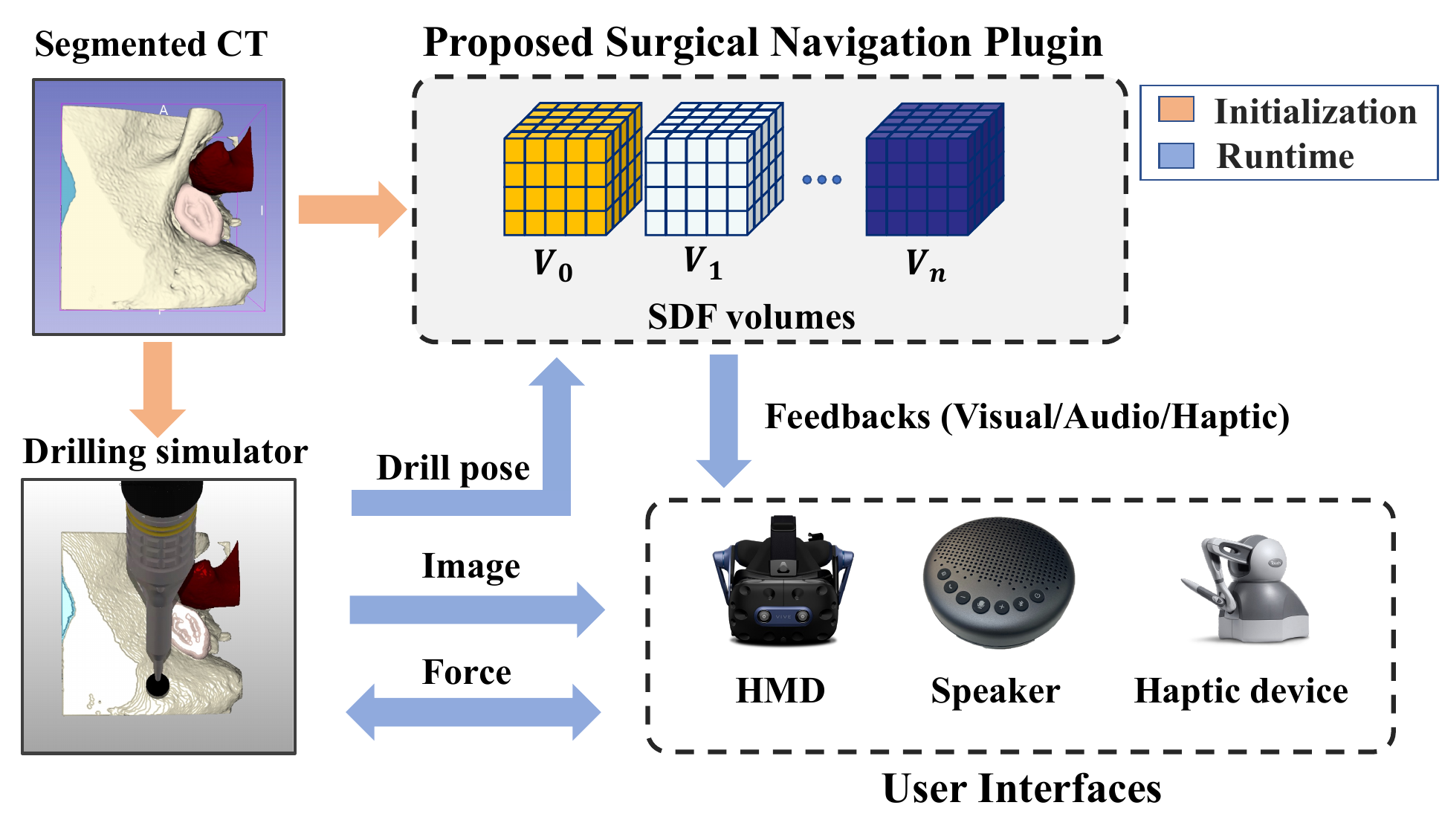}
      
      \captionof{figure}{System Architecture. The proposed Surgical navigation plugin is developed on top of the FIVRS framework\cite{FIVRS2023}. SDF calculation is done at the initialization phase using the same CT scan that is loaded into the simulation environment. At runtime, multimodal feedback is provided to the users via the haptic device, speakers, and head mounted display incorporated in the FIVRS system.}
      \label{fig:system_overview}
 \end{figure*}

Surgical simulation environments present a cost-effective solution to the problems of designing and evaluating guidance systems. First, surgical simulation allows testing the navigation systems on highly realistic surgical environments using
anatomic models created from CT or MRI images. Second, simulations are controlled environments that are well suited to evaluate the effect of guidance on surgical situational awareness, perceived workload, and skill maintenance.  Nevertheless, adding surgical navigation capabilities to existing simulation environments is still a challenging task, requiring the integration of multiple software components without degrading the simulation performance. For example, providing the surgeon with a real-time warning when a surgical tool is getting close to delicate anatomy requires very efficient calculation of tool-to-model distances.

To facilitate the development process of novel guidance systems, we have developed a modular and open-source plugin that enables multimodal surgical navigation for the Asynchronous Multi-Body Framework (AMBF) \cite{AMBF_main_paper} simulator (Fig. \ref{fig: Exp_setup}). At the heart of our plugin, we have integrated a library to calculate Signed Distance Fields (SDF) from segmented anatomical volumes at initialization. The resulting SDF volumes can then be used to calculate in real-time the minimum distance between the virtual instruments and different anatomies and provide feedback to the user (Fig. \ref{fig:system_overview}).

The plugin was designed to be highly applicable to various surgical procedures that rely on imaging data. The flexibility of this framework can be attributed to two reasons. Firstly, the plugin directly supports the loading of segmented anatomical volumes created with 3D Slicer \cite{kikinis20143d}, an open-source platform popular among clinical users for analyzing and displaying information derived from medical imaging. Secondly, the plugin was designed to allow easy customization of the feedback modalities, allowing it to be used on multiple procedures and surgical specialties with potentially different safety constraints.

Using this framework, we have developed a multimodal surgical navigation system for the skull base surgery simulation environment FIVRS \cite{FIVRS2023}. 
We have two goals for the navigation system: (1) identifying skull base surgeons' preferred feedback modalities to navigate anatomical variability; and (2) demonstrating the flexibility of our proposed method to account for different types of feedback modalities.  The selected feedback modalities for this system were visual, auditory, and haptic. A pilot user study was conducted with three experienced surgeons and a medical student to evaluate the utility of the system and guidance modalities. The results of this pilot study will be used in the future to guide a larger user study aimed at identifying the optimal feedback modalities in skull base surgery.

%The effectiveness of each type of guidance was assessed using a combination of a subjective workload assessment questionnaire and objective performance metrics, such as completion time and the number of collisions with critical anatomies. \

\begin{figure*}[ht]
    \centering
    \includegraphics[width=1.0\textwidth]{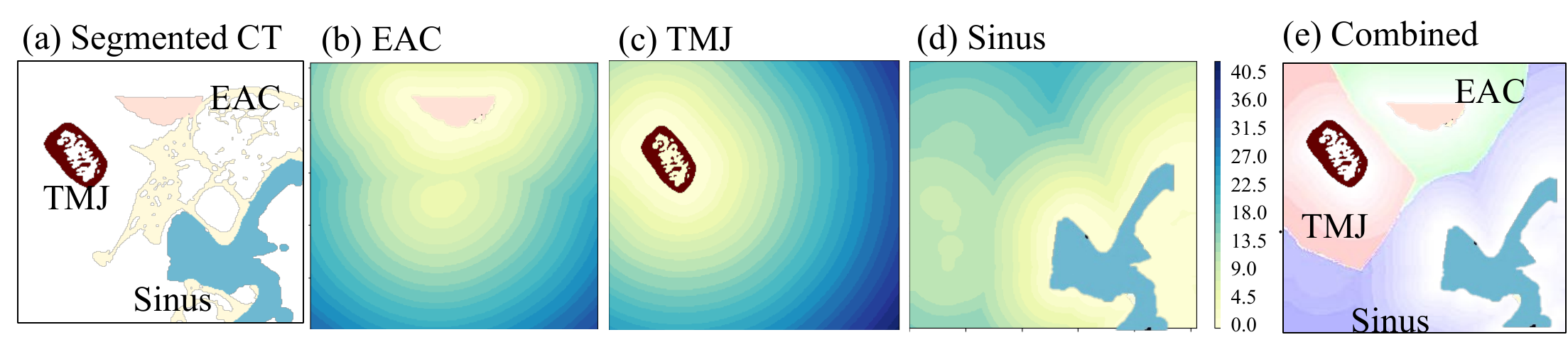}
    \caption{Example visualization of an SDF volume's slice. (a) Segmented CT scan showing three anatomies: Temporomandibular (TMJ), Ear Canal (EAC), and Sinus, (b) SDF slice for \hi{EAC}, (c) SDF slice for \hi{TMJ}, and (d) SDF slice for Sinus. Voxels at each slice store the minimum distance between that voxel's location and a specific anatomy. The units for the color scale are $mm$. (e) Combined SDF image of TMJ, EAC, and Sinus. In the combined slice, different regions are color-coded by the closest anatomy (Green: EAC, Red: TMJ, Blue: Sinus).}
    \label{fig:EDT_diagram}
\end{figure*}
%Regions are color coded differently according to the closest anatomy. 
 
Although the proposed framework was evaluated on a single surgical procedure, we emphasize that our plugin can be used to develop surgical navigation systems in any surgical specialty that relies on CT scans, e.g., sinus, orthopedic, spinal, and laryngeal surgery. In summary, this paper reports the following contributions: (1) a multimodal navigation system for skull base surgery; (2) an open-source and modular library that enables the development of IGSN systems based on Signed Distance Field for a wide range of surgical procedures; and (3) a pilot study showing the utility of the system in the context of a mastoidectomy procedure.

\section{Related Work}

Traditionally, image-guidance systems have relied on additional screens to display the position of surgical tools relative to patient anatomy. However, these systems are rarely used in temporal bone surgery as the surgeon would have to switch their attention from the surgical field to the guidance screen \cite{schwam_utility_2020}. In this regard, less intrusive feedback modalities such as audio, visual overlays, and haptic feedback have been shown to be more promising in mastoidectomy procedures. An audio guidance system was proposed by Cho \textit{et al.} \cite{cho_warning_2013} to avoid damage to the facial nerve. This system would gradually increase the alarm frequency as the drill got closer to the optical nerve to alert the surgeon.

Regarding visual guidance, one common approach has been to use a head mounted display (HMD) to annotate the surgeon's field of view. For example, Rose \textit{et al.} \cite{rose_development_2019} used a Microsoft HoloLens HMD to overlay transparent images of the neck and temporal bone anatomy on phantom models. Finally, haptic feedback provided by cooperative-control robots has been proposed as a mechanism to improve safety in surgery. Ding \textit{et al.} \cite{ding_volumetric_2021} demonstrated that virtual fixtures could be enforced by 
a cooperatively controlled robotic system with sub-millimeter accuracy in a phantom drilling experiment. 

Our current system integrates multiple previously proposed feedback modalities into a single system and uses SDF information to activate them in a timely manner. This allows surgeons to experience multiple types of feedback on the same simulated task, enabling objective comparisons across the modalities. Furthermore, testing the guidance modalities in a repeatable and controllable environment, such as a VR simulation, isolates the effects of different modalities on performance and mental demand.

\section{Methodology}
% We developed an AMBF plugin that provides real-time guidance for surgical training of any hard tissue surgery using IGSN. To achieve navigation in real-time, Signed Distance-Field (SDF) volumes are calculated offline and loaded to the plugin. After loading the volumes, the distance between the virtual instrument and the closest anatomical structure is easily obtained by accessing all the SDF volumes at a specific location and selecting the smallest value. Utilizing the proposed plugin, a multimodal navigation system is presented. 
The development of our multimodal surgical guidance system is presented as follows.  In section \ref{sec_sdf_df}, the definition of SDF volumes and the format used to represent the anatomy is presented. In section \ref{sec_sdf_cal}, the definition of an SDF volume and the library used to calculate them is presented. Finally, section \ref{sec_feedback_modalities} describes the development of three SDF-based feedback modalities.  

\subsection{Anatomy representation and SDF definition} \label{sec_sdf_df}

We use 3D Slicer to visualize and segment preoperative CT and MRI images. For our study, we 
use patient models derived from a collection of patient CT scans.  Multiple anatomic structures for each patient were segmented using methods developed by 
Ding \textit{et al.} and saved in Segmented Nearly Raw Raster Data (``.seg.nrrd'') format \cite{ding2022automated}. This data is then loaded in the simulation and rendered in the scene as described in \cite{FIVRS2023}.

The same segmented CT images are used for SDF generation to ensure consistency between the SDF's voxel coordinates and the surgical simulation models. A plugin was developed to load the SDF volumes at initialization and to query the stored values at runtime. 

\begin{figure*}[ht]
    \centering
    \includegraphics[width=1.0\textwidth]{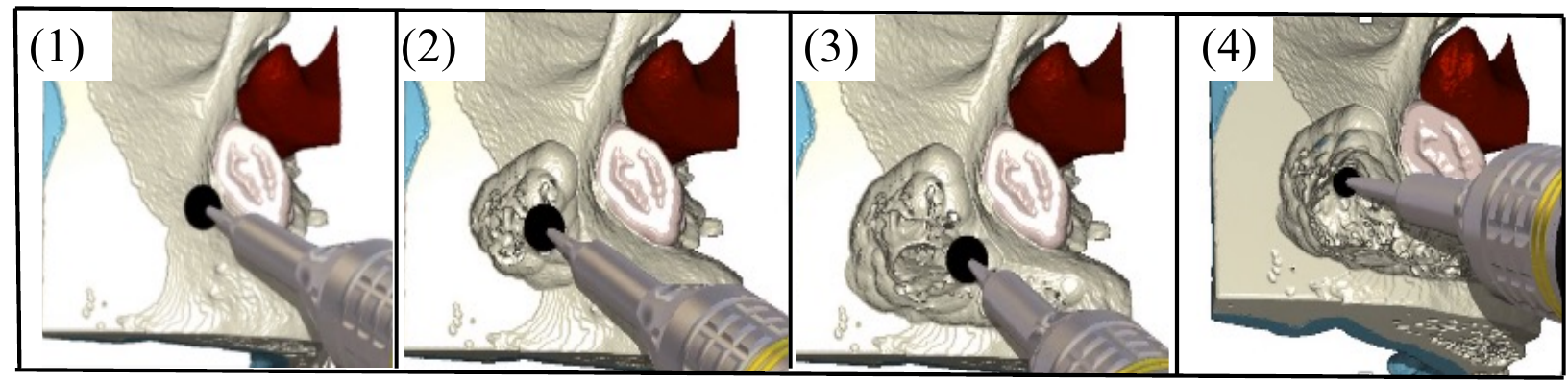}
    \caption{Sequence of snapshots (from 1 to 4) showing the Mastoidectomy procedure performed by the user study participants. }
    \label{fig:mastoidectomy-snapshots}
\end{figure*}

\subsection{Calculation of SDF} \label{sec_sdf_cal}
Each segmented CT scan comprises of several anatomies ($n=16$ in our current experiments). To calculate the SDF volume from the $n^{th}$ anatomy, $S^{(n)}$, a C++ open source library was used \cite{KazhdanSDFLibrary}. This library provides an efficient and parallelized implementation of Saito and Toriwaki's  method \cite{saito_new_1994} for SDF calculation. 
An SDF volume is represented as a 3D voxel grid where the value at each voxel represents the signed distance to the closest point on a specific anatomy (Fig. \ref{fig:EDT_diagram}). The positive and negative values of the distance indicate voxels that are exterior and interior to the anatomy, respectively. 

\subsection{Feedback modalities based on SDF volumes} \label{sec_feedback_modalities}

The loaded SDF volumes allow the system to easily query the distance to the closest anatomy. Using the distance to the closest anatomy, three distinct feedback modalities were developed to improve the user's situational awareness: visual,  haptic, and auditory feedback.
For all modalities, the drill's position is converted to the SDF frame giving us the corresponding voxel coordinate ($x\equiv \{i, j, k\}$) within the SDF volumes. The minimum distance between the drill tip and the nearest anatomy is calculated by querying all the SDF volumes ($S^*(x) \leq \forall S^{(n)}(x)$). The SDF volume for the closest anatomy ($S^*(x)$) is used to generate user feedback. Haptic and audio feedback is activated once the drill is within the predefined thresholds. These thresholds were selected by an expert otolaryngology surgeon.

\subsubsection{Visual Feedback}

\begin{figure}[hb]
\centering
    \includegraphics[width=0.4\textwidth]{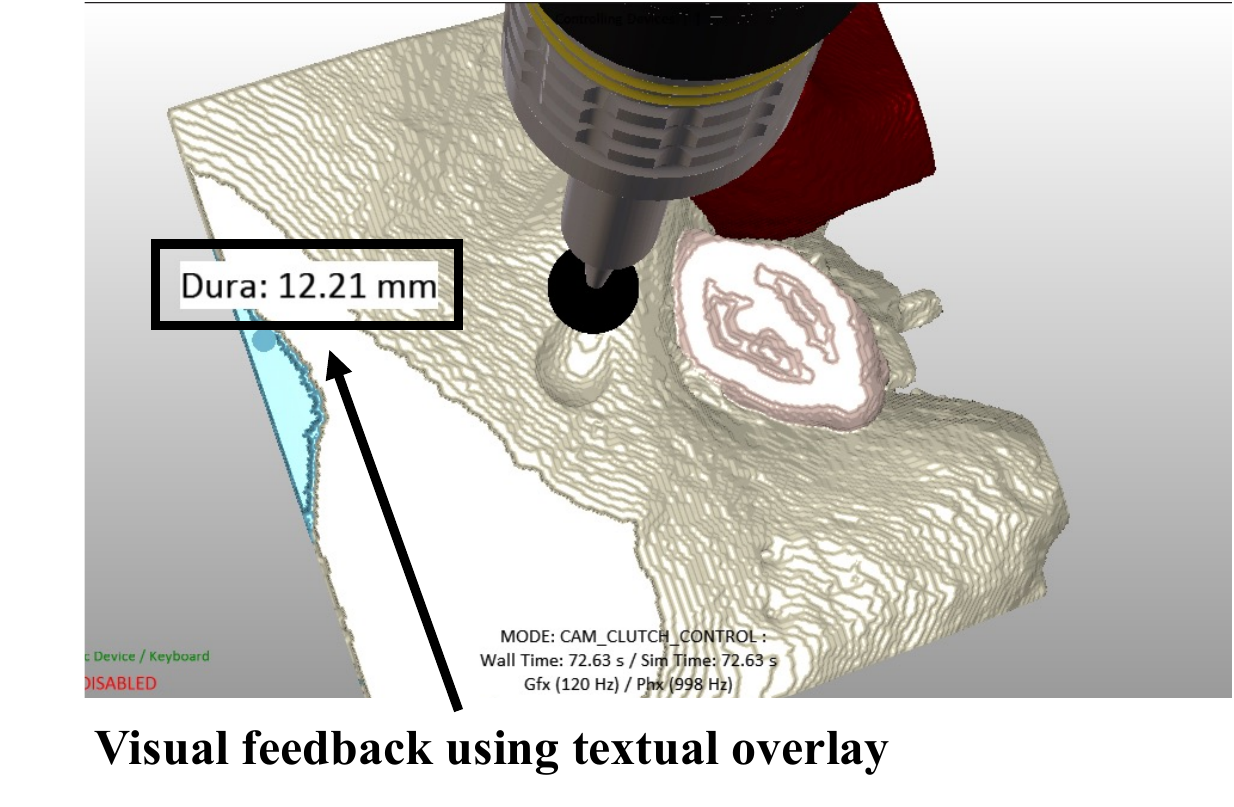}
     \caption{Visual Feedback. Textual overlay provides the name of the closest anatomy and the distance to that specific anatomy via HMD in stereoscopic view.}
     \label{fig:Visual1}
\end{figure}

Our method uses the SDF to display the closest anatomy's name and the distance to it in a text box on the head-mounted stereoscopic display (Fig. \ref{fig:Visual1}). The text changes its color once the drill is closer than $1mm$ to any anatomy. The location of this text box was also determined after a pilot study with an attending Otologic surgeon.    

\subsubsection{Audio Feedback}
We implement auditory feedback to notify the surgeons when the tool tip is about to collide with critical anatomies. Otologic surgeons are familiar with this kind of auditory feedback from nerve proximity monitors (e.g., \cite{mangia2020}), and it provides a form of situational awareness that is not disruptive during the surgery. An alarm sound is generated when the drill is closer than a defined distance $\tau_{a}$ to the critical anatomies. Furthermore, after consulting with surgeons, we discovered that audio feedback could provide initial situational awareness cues when approaching the critical anatomy so that the activation threshold for audio warnings can be larger than that used for haptic feedback.

\subsubsection{Haptic Feedback}
FIVRS adopts CHAI3D’s \cite{conti2005chai} finger proxy collision algorithm \cite{ruspini2001haptic} to provide haptic feedback by simulating the collision of the drill tip with the surface of the volume.
% The movement of the tool is simulated using the proxy/goal spheres such that the proxy sphere cannot cross the boundary of the surface or volume while the goal sphere will move freely to match the tool.
% Force feedback is generated proportional to the distance between the proxy and the goal spheres.
We add an additional force term, $F_{SDF}\in R^3$, to the contact force provided by FIVRS to prevent the user from drilling critical anatomies. The formulation of this SDF-based force can be written as follows:

\begin{equation}
F_{SDF} =
    \begin{cases}
        F_{max}(\tau_{f} - d_{a} ){\vec{d}^{\ (SDF)}} & \text{if } d_{a} < \tau_{f}\\
        0 & \text{Otherwise}
    \end{cases}
\end{equation}

where $F_{max} \in R$ is the maximum force in Newtons, $d_{a}$ represents the closest distance to the anatomy,  $\tau_{f}$ is the activation threshold for haptic feedback and $\vec{d}^{\ (SDF)}$ the direction of the force. This direction is calculated with $\vec{d}^{\ (SDF)}= \vec{d}/|\vec{d}|$, where $\vec{d}$ is a finite difference.

\begin{figure*}[ht]
     \includegraphics[width=0.5\textwidth]{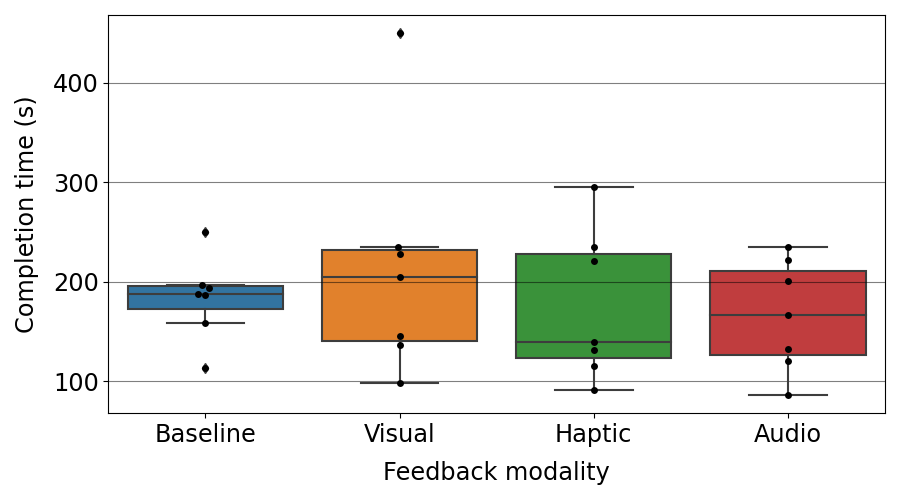}
     \includegraphics[width=0.5\textwidth]{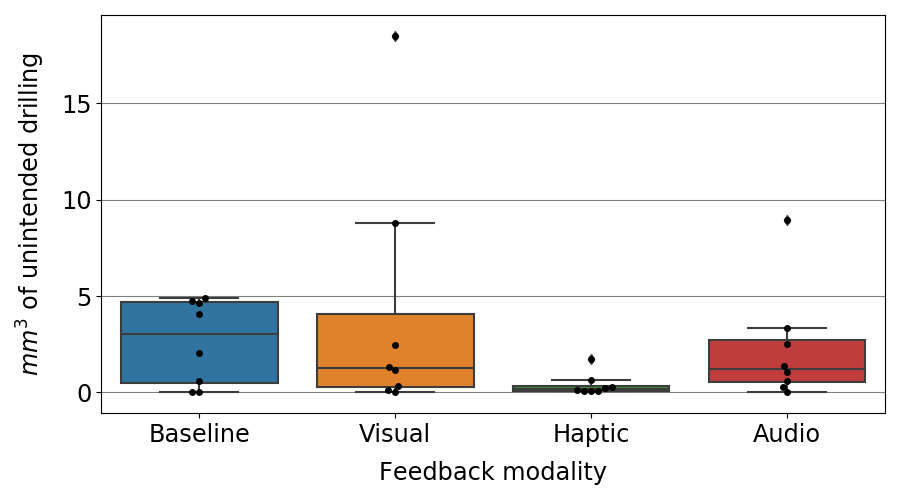}
    \caption{User study objective metrics. (a) Completion time per anatomy and (b) number of unintended voxels removed. \hi{Black dots in the figure represent the data points used to construct the boxplots. Data points are separated horizontally for clarity.}}
    \label{fig:metrics_all}
\end{figure*}

\section{Evaluation and results}\label{sec2}
We conducted a pilot user study to illustrate the use of our framework and to provide preliminary feedback on the proposed guidance modalities for a larger follow-on study. For the experimental setup, we employed a Phantom Omni (3D Systems, USA) as a haptic device and a VIVE PRO (HTC VIVE, Taiwan) as a Head Mounted Display. The entire system was installed on a single transportable workstation (Fig. \ref{fig: Exp_setup}).

\subsection{User Study Design}
Four clinical participants (2 
attending surgeons, 1 fellow, and 1 medical student) who all have sufficient knowledge about mastoidectomy were recruited in this study. We selected two segmented CT scans of the temporal bone, both of which contain the same 16 distinctive anatomies. Based on expert surgeon feedback, we confirmed that these two CT scans require similar skills to perform the mastoidectomy; thus, we assume that there is no significant difference in the complexity of the procedure.
Each user was asked to drill a wide cortical mastoidectomy to the point of exposing the short process of the incus without harming the non-bone anatomies (Fig. \ref{fig:mastoidectomy-snapshots}). We tested four different feedback modalities: (1) No assistance, (2) with visual feedback, (3) with audio feedback, and (4) with force feedback. \hi{For every condition, users experience the contact force implemented in FIVRS, and for condition (4), the proposed haptic feedback is enforced on top of the contact force.} Experimental order and the drilled anatomy were randomized to mitigate the learning effect. Lastly, each user was allowed to familiarize themselves with the simulator before starting the experiment.

To evaluate the proposed modalities, we adopted two objective metrics, which are the task completion time and the number of unintended voxels removed, and one subjective metric that is the NASA TLX survey. Task completion time was defined as the time between the first and last removed voxel. The number of removed voxels from critical anatomy was also recorded during the experiment.

After each trial, users were asked to complete the NASA TLX \cite{hart2006nasa} survey. This survey uses six indicators to evaluate the workload: \textit{mental demand, physical demand, temporal demand, performance, effort, and frustration}. 

\begin{table}[ht]
\caption{\hi{Objective performance metrics results.}}
\label{table:results}

% \resizebox{\columnwidth}{!}{%
\begin{tabular}{|cccccc|}
\hline
\multicolumn{6}{|c|}{\textbf{Pilot study objective performance metrics}}                                                                                                                  \\ \hline
\multicolumn{2}{|c}{\textbf{Modality}}                                                                           & \textbf{Baseline} & \textbf{Visual} & \textbf{Haptic} & \textbf{Audio} \\ \hline
\multicolumn{1}{|c|}{}                                        & \multicolumn{1}{c|}{Mean}                        & 182               & 214             & 175             & 173            \\
\multicolumn{1}{|c|}{\multirow{-2}{*}{Time (s)}}              & \multicolumn{1}{c|}{\cellcolor[HTML]{FFFFFF}Std} & 38.5              & 107             & 69.4            & 55.6           \\ \hline
\multicolumn{1}{|c|}{Unintended voxels}                       & \multicolumn{1}{c|}{Mean}                        & 375               & 584             & \textbf{57.8}   & 325            \\
\multicolumn{1}{|c|}{\cellcolor[HTML]{FFFFFF}removed (count)} & \multicolumn{1}{c|}{Std}                         & 312               & 927             & \textbf{82.6}   & 417            \\ \hline
\end{tabular}
% }
\\\\
\textit{\hi{Among all experimental conditions, haptic guidance led to a reduction of unintended voxels removed without increases in completion time.}}
\end{table}

% \begin{table}[ht]
% \fontsize{9}{9}\selectfont
% \begin{tabular}{c|c| c c c c c}
%      & & Baseline & Visual & Haptic & Audio \\
%      \hline \rule{0pt}{2ex}  
%               & Max & 250  & 450 & 295  & \textbf{235}\\
%               & Min & 113  & 98.4& 91.8 & \textbf{85.8}\\
%      Time (s) & Mean& 182  & 214 & 175  & \textbf{173}\\
%               & Std & \textbf{38.5} & 107 & 69.4 & 55.6\\
%               \hline \hline \rule{0pt}{2ex}  
%     Unintended& Max  & 696 & 2641& \textbf{250}  & 1277 \\
%     Voxel     & Min  & 8   &\textbf{ 0}   & \textbf{0}    & \textbf{0} \\
%     Removed   & Mean & 375 & 584 & \textbf{57.8} & 325\\
%     (count)   & Std  & 312 & 927 & \textbf{82.6} & 417\\  
% \end{tabular}
% \caption{Results from the pilot user study.}
% \label{table:results}
% \end{table}

\subsection{Results}

The task completion times and the number of unintended voxels removed are shown in Table \ref{table:results} and Fig. \ref{fig:metrics_all}.
In terms of task completion time, there was no significant difference between the baseline and the three proposed assistance methods. Fig. \ref{fig:metrics_all} shows that haptic feedback reduced the number of inadvertent voxels removed to nearly zero. Audio feedback also reduced the number of unintended voxels removed.

\begin{figure}[ht]
    \centering
    \includegraphics[width=0.55\textwidth]{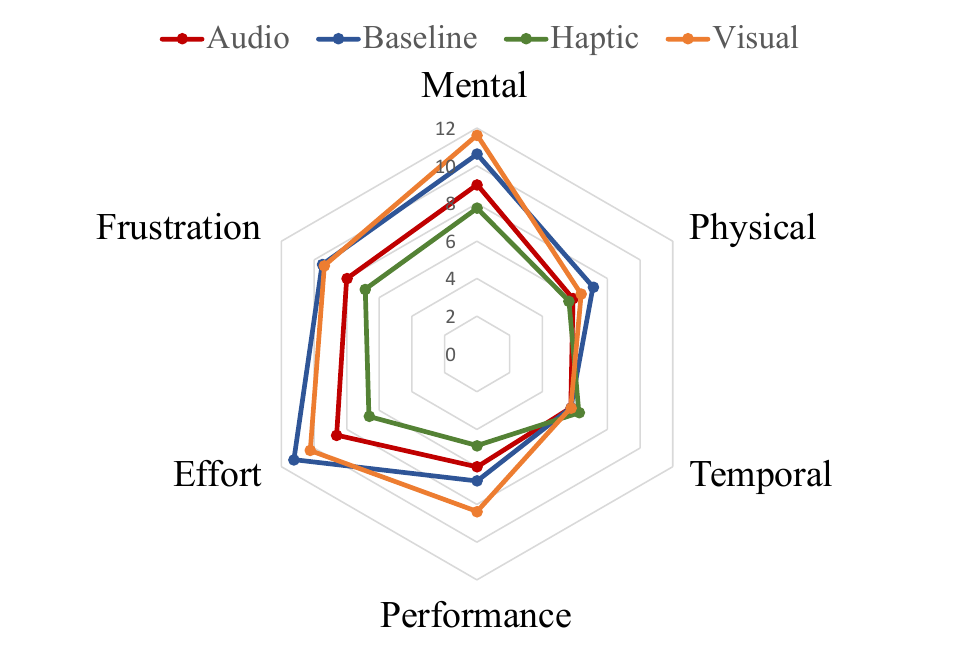}
    \caption{NASA TLX results. Values near the center indicate better results.}
    \label{fig:result_nasa}
\end{figure}

The result of the NASA TLX survey can be found in Fig. \ref{fig:result_nasa}, 
which shows that haptic feedback has the lowest workload in \textit{mental demand, physical demand, performance, effort, and frustration} across all four conditions. Audio feedback also reduced most of the workload (\textit{mental demand, physical demand, performance, effort, and frustration}) compared to the baseline method. On the other hand, visual feedback led to higher \textit{mental demand, performance} compared to baseline.

\section{Discussion}\label{sec12}
During the user study, the proposed SDF plugin successfully offered three different assistance modalities at an interactive update rate of over 80 $Hz$ on an AMD Ryzen 5800 CPU, 32 GB DDR4 RAM, and RTX 3080 GPU system. This suggests that the proposed pipeline was able to generate real-time guidance from patients' CT scans.
The user study results showed that haptic feedback was able to reduce the number of unintended contacts with critical anatomies and improve multiple workload metrics. Although audio feedback did not impose physical restrictions to avoid collisions, most participants felt that it was useful and reduced mental demands. One of the participants did note, however, that not knowing which anatomy triggered the feedback increased his mental workload.

Visual feedback had a negative impact on drilling assistance for most participants, which can be attributed to the type of information we offered. One of the participants claimed that the location of the text overlay was too distant from the workspace and showing the distance was too much information to process; consequently, they had to shift their view from the drill. These findings align with the results indicating that visual feedback has a high variance in the unintended voxels removed, as well as the NASA TLX results. To address these issues, an alternate visual feedback method will be implemented in future work in which the distance information is provided by changing the drill or anatomy's color. Lastly, the less experienced participant found the visual feedback to be a helpful learning tool since it provides information about anatomies that are not physically visible.

%An alternate visual feedback method to tackle this issue is to represent the distance by changing or overlaying the color of the drill burr or the color of the anatomy. This method is seemingly more intuitive, but implementing the same functionality in a cooperatively controlled robot is challenging due to registration issues and the difficulty of overlaying augmented reality graphics in the operating microscopes commonly used in these procedures. 

\section{Summary and Future Work}
\label{sec13}
This paper reports the development of a novel and highly applicable framework to develop real-time surgical navigation systems based on the Signed Distance Field (SDF). \hi{The framework was designed to be adaptable to any surgical procedure that deals with rigid structures, relies on CT or MRI imaging and could benefit from multiple feedback modalities.} Using our proposed method, we developed a multimodal guidance system (visual, audio, and haptic) for a mastoidectomy virtual drilling simulator. A pilot user study with 3 surgeons and 1 medical student showed that our guidance system reduced unintended contact with critical anatomies and lowered mental demands without increasing operation time. We also found that users preferred haptic and audio feedback over visual feedback.  
% The framework was designed to be adaptable to any surgical procedure that relies on CT or MRI imaging and to account for different feedback modalities.
In future work, we plan to conduct more extensive studies analyzing the effect of multiple feedback modalities on surgical performance. Furthermore, we plan to develop guidance systems for different surgical procedures, such as laminectomy and sinus surgery, to test the general applicability of our system. Lastly, combining the digital twins framework \cite{shu2022twin}, we aspire to implement SDF-based guidance modalities with existing robotic systems and real surgical phantoms.

\section*{Acknowledgments and Disclosures}
 Hisashi Ishida was supported in part by the ITO foundation for international education exchange, Japan.  
 This work was also supported in part by a research contract from Galen Robotics, by NIDCD K08 Grant DC019708, by a research agreement with the Hong Kong Multi-Scale Medical Robotics Centre,
and by Johns Hopkins University internal funds.
Russell Taylor and Johns Hopkins University (JHU) may be entitled to royalty payments related to the technology discussed in this paper, and Dr. Taylor has received or may receive some portion of these royalties. Also, Dr. Taylor is a paid consultant to and owns equity in Galen Robotics, Inc.  These arrangements have been reviewed and approved by JHU in accordance with its conflict of interest policy.  
This study was approved by the Johns Hopkins University Institutional Review Board under the protocol IRB00264318.

We thank Dr. Michael Kazhdan for the assistance with SDF implementation.

\section*{Supplementary information} 
A supplementary video is provided with the submission. For more information, visit the project repository at \url{https://github.com/hisashiishida/SDF_based_assistance}.

\bibliography{IROS2023-SDF}
\bibliographystyle{IEEEtran}

\end{document}